\documentclass[12pt]{article}
\usepackage{graphics}
\begin{document}
\begin{center}
{The Self-energy of Nucleon for the Pole term of the Axial-vector Current and the Neutron $\beta$-decay}
\end{center}
\begin{center}
{Susumu Kinpara}
\end{center}
\begin{center}
{\it Institute for Quantum Medical Science (QST) \\ Chiba 263-8555, Japan}
\end{center}
\begin{abstract}
The effect of the pion propagator on the $\beta$-decay of neutron is investigated to account for
the ratio of the axial-vector to the vector part $g_A/g_V$ by including the pole term.
A suggestion is made that the sign and the size of the vertex correction is appropriate to improve the discrepancy
between the result of the model hamiltonian and the experiment.
\end{abstract}
\section*{\normalsize{1 \quad Introduction}}
\hspace*{4.mm}
For the investigation of the mechanism of the reaction processes the nucleon is the elementary object in the intermediate energy regions.
The idea of the point-like structure is inevitably modified by the interaction with the other particle as a probe.
Particularly the lightest meson that is pion plays a decisive role on improvement of the state of nucleon.
It is expressed by the propagator and the effect of the pion degrees of freedom is essential to evaluate the constants related to each process. 
\\\hspace*{4.mm}
The self-energy of nucleon is the starting point of the pion-nucleon system and also changes the form of the vertex part of the interaction.
The relation between these functions is obtainable by using the equation of motion independent of the calculation of the perturbative expansion.
Because of the derivative coupling the pseudovector $\pi$-N interaction appears to be difficult to control the divergences 
in the framework of the renormalization by the counter terms of the mass and the fields.
In turn, by virtue of the non-perturbative term stemmed from the non-perturbative relation the divergences are removed together with the counter terms 
mentioned above.
\\\hspace*{4.mm}
Adding the non-perturbative term the pseudovector coupling is connected to the pseudoscalar coupling.
The additional interaction contains the self-energy and the vertex part is modified by the term.
When the calculations of the pseudoscalar $\pi$-N system in free space do not give the perfect results
the term of the self-energy is expected to account for the remaining gap. 
\\\hspace*{4.mm}
The method to construct the self-energy is based on the field theoretical treatment giving the form of the function as the series of powers in the Dirac operator.
Applying it to the scattering processes the structure constants such as the magnetic moment and the polarizability of nucleon
and the $\pi$-N scattering parameters of the $S$ and $P$ waves are described well by each model chosen separately for the respective phenomenon.
\\\hspace*{4.mm}
For the low energy regions below the $\Delta$(1232) resonance the lowest-order approximation of the perturbative calculation with the self-energy 
does not attain to describe the $\pi$-N elastic scattering.
The higher-order calculations could change the form of the self-energy largely so as to provide the volume of the low energy parameters well.
On the other hand the method of the matrix inversion enables us reach the set of the scattering parameters which construct the self-energy as the output. 
The application of it to the photoproduction of pion is interesting for the pseudoscalar model needs the additional effects
from the threshold to the resonance energy.
In the present study the $\beta$ decay of neutron is chosen to examine the form of the self-energy by the meson-exchange model of the field theory.
\\
\section*{\normalsize{2 \quad The vertex function with the axial-vector current}}
\hspace*{4.mm}
The hadronic part of the lagrangian of the interaction $L_{int}$ is assumed to consist of the vector $L_v$ and the axial-vector $L_a$ parts
\begin{eqnarray}
L_{int} = L_v + L_a = -g \, \bar{\psi}(x) \,\gamma_\mu \,(1-\gamma_5) \,\vec{\tau} \,\psi(x) \,\vec{V}^\mu
\end{eqnarray}
in which the $\psi$ and $\vec{V}_\mu$ are the fields of nucleon and the isovector vector boson respectively.
The strength of the coupling constant $g$ is assumed to be comparable with the electric charge and then allows us to expand the term in powers of $g$
and neglect the terms of the higher-order practically.
The $\vec{\tau}$ is the 2$\times 2$ isospin matrix.
\\\hspace*{4.mm}
In order to derive the non-perturbative relation and examine the structure of the vertex part which unites the external lines of some bosons and fermions
the set of two isovector currents that is the vector current $\vec{B}_\mu$ and the axial-vector current $\vec{A}_\mu$ 
\begin{eqnarray}
(\vec{B}_\mu,\vec{A}_\mu) \equiv \bar{\psi}\,(\gamma_\mu,\gamma_5\gamma_\mu)\,\vec{\tau}\,\psi
\end{eqnarray}
is indispensable.
The sum $\vec{J} \equiv \vec{B}+\vec{A}$ suffices the relation
\begin{eqnarray}
\partial \cdot \vec{J} = - 2 i M \vec{\rho}_5 + 4 \,g \, \vec{V} \times \vec{J}
\end{eqnarray}
\begin{eqnarray}
\vec{\rho}_5 \equiv \bar{\psi}\,\gamma_5\,\vec{\tau}\,\psi
\end{eqnarray}
where the nucleon mass $M$ is the average value of proton and neutron. 
\\\hspace*{4.mm}
The equation of motion for $\vec{V}$ is given as
\begin{eqnarray}
(\partial^2 + m^2) \vec{V} = g \, (1+m^{-2} \,\partial \partial\cdot)\vec{J} 
\end{eqnarray}
in which the weight of the mass of the boson $m$ is not definite and possibly simulate the $\rho$-meson 
or the much heavior boson related to the weak interaction.
In practice the $O(g^2)$ term is dropped approximately to derive the relations below. 
\\\hspace*{4.mm}
The generalized relation is derived for the $T$-product which consists of a set of the field operators 
\begin{eqnarray}
T[\,V_\mu^i(x) \cdots \, \varphi_a (x_a) \cdots]
\end{eqnarray}
in which $\varphi_a(x_a)$ ($\,\equiv \, \psi(x_a)$ or $\bar{\psi}^{\small T}(x_a)\,$) represents the Dirac fields.
Operating $\partial^2+m^2$ on the quantity (6) and using Eq. (5) without the $O(g^2)$ term and the following relation
\begin{eqnarray}
\delta(x_{a0}-x_0)\,[\varphi_a(x_a),\,\rho_5^i(x)]=e_{ai}\varphi_a(x_a)\,\delta^4(x_a-x)
\end{eqnarray}
with $e_{ai}=\gamma_0 \gamma_5 \tau_i$ for $\varphi_a(x_a)=\psi(x_a)$ or $(\gamma_0 \gamma_5 \tau_i)^T$ for $\varphi_a(x_a)=\bar{\psi}^{\small T}(x_a)$
the generalized form of the equation of motion is obtained  
\begin{eqnarray}
(\partial^2 + m^2) T[\,V_\mu^i(x) \cdots ] 
= g\, T[\, J_\mu^i (x) \cdots \,] +i \frac{\delta}{\delta V^\mu_i(x)}T[\,\cdots\,] \nonumber\\\nonumber\\
-2 i g m^{-2} M (\,\partial_\mu T[\rho_5^i(x)\cdots]+g_\mu^0\sum_{a,b,\cdots}e_{ai}\delta(x-x_a)T[\cdots]\,)
\end{eqnarray}
The third term of the right-hand side comes from the breaking of the conservation of the axial-vector current 
in comparison with the case of the quantum electrodynamics (QED).
Another main difference from the QED is the part of the functional derivative in Eq. (8) defined as
\begin{eqnarray}
\frac{\delta V_\nu^j(x^\prime)}{\delta V^\mu_i(x)} = \delta_{ij} \,\rho_{\mu\nu}(x-x^\prime)
\end{eqnarray}
\begin{eqnarray}
\rho_{\mu\nu}(x-x^\prime) = \int \frac{d^4 k}{(2 \pi)^4} \, \rho_{\mu\nu}(k) \, e^{-k \cdot (x-x^\prime)}
\end{eqnarray}
\begin{eqnarray}
\rho_{\mu\nu}(k) = g_{\mu\nu} - \frac{k_\mu k_\nu}{m^2} - (1-\frac{k^2}{m^2})\eta_\mu \eta_\nu
\end{eqnarray}
where $\eta_\mu = (1,0,0,0)$.
The additional terms in Eq. (11) are inherent in the massive vector boson.
The delta part ($\sim (m^2-k^2)\eta_\mu \eta_\nu$) is dropped due to the cancellation with the normal dependent term of the interacting hamiltonian. 
\\\hspace*{4.mm}
In spite of the breaking of the current conservation the differentiation of Eq. (8) and the use of the relation of the commutator
\begin{eqnarray}
\delta(x_{a0}-x_0)[\varphi_a(x_a),J_{0i}(x)] = P_{ai}\,\varphi_a(x_a)\,\delta^4(x_{a}-x)
\end{eqnarray}
with $P_{ai}=(1-\gamma_5) \tau_i$ for $\varphi_a(x_a)=\psi(x_a)$ or $-(1+\gamma_5) \tau_i^{\small T}$ for $\varphi_a(x_a)=\bar{\psi}^{\small T}(x_a)$
gives the useful form
\begin{eqnarray}
&&(\partial^2 + m^2) \partial^\mu \, T[\,V_\mu^i(x) \cdots ] 
= -2 i g m^{-2} M (\partial^2+m^2)\, T[\, \rho_{5i} (x) \cdots \,] \nonumber\\\nonumber\\ 
&&+[\,i \partial^\mu \frac{\delta}{\delta V^\mu_i(x)}
-g\sum_{a,b,\cdots} (P_{a i}+2 i m^{-2} M e_{ai} \,g_\mu^0 \,\partial^\mu)\,\delta(x-x_a)\,]\,T[\cdots]
\end{eqnarray}
in which the approximate form of the current $\partial \cdot J_i(x) \approx -2 i M \rho_5^i(x)$ is used.
\\\hspace*{4.mm}
When the $T$-product consists of two boson fields the general relation in Eq. (13) gives the equation for the boson propagator 
\begin{eqnarray}
i D^{ij}_{\mu\nu}(x-y) = \langle\, T[V_\mu^i(x) \, V_\nu^j(y)] \,\rangle
\end{eqnarray}
\begin{eqnarray}
(\partial^2 + m^2)\partial^\mu D^{ij}_{\mu\nu}(x-y) = \partial^\mu \rho_{\mu\nu}^{ij}(x-y) 
\nonumber\\\nonumber\\ 
-2 g m^{-2} M (\partial^2 + m^2) \langle\, T[\rho_5^i(x) \, V_\nu^j(y)] \,\rangle 
\end{eqnarray}
\hspace*{4.mm}
Using the $T$-product of three fields the vertex function $\Gamma_i^\mu(x\,y\,;z)$
is defined so as to give the following relation
\begin{eqnarray}
&&\langle\, T[\psi(x) \, \bar{\psi}(y) \, V_\mu^i(z)] \,\rangle  \nonumber\\
&&\equiv -g \int d x^\prime d y^\prime d z^\prime \, G(x-x^\prime) \, \Gamma_j^\nu(x^\prime\,y^\prime\,;z^\prime) \, G(y^\prime-y) \, D_{\nu\mu}^{ji}(z^\prime-z)
\end{eqnarray}
with the nucleon propagator
\begin{eqnarray}
i G(x-y) = \langle\, T[\psi(x) \, \psi(y)] \,\rangle
\end{eqnarray}
Operating $(\partial_z^2 + m^2)\partial_z^\mu$ on Eq. (16) with respect to $z$ and applying Eq. (13) for three fields and Eq. (15) 
the relation between these functions are given as follows
\begin{eqnarray}
&&\int d x^\prime d y^\prime d z^\prime \, \rho_{\mu\nu}^{ij} (z-z^\prime) \, G(x-x^\prime) \, \partial_{z^\prime}^\mu \Gamma_j^\nu(x^\prime\,y^\prime\,;z^\prime)
\, G(y^\prime-y) \qquad\qquad\qquad\nonumber\\\nonumber\\
&&= i \delta(z-x) \,(1-\gamma_5) \,\tau_i \,G(x-y) - i \delta(z-y) \,G(x-y) \,(1+\gamma_5) \,\tau_i \nonumber\\\nonumber\\
&& +2 i m^{-2} M (\partial_z^2 + m^2) [ \,\langle\, T[\,\rho_{5 i}(z) \psi(x) \, \bar{\psi}(y) \,] \,\rangle \nonumber\\\nonumber\\
&& -i g \int d x^\prime d y^\prime d z^\prime \,\langle\, T[\,\rho_{5 i}(z) V_\nu^j (z^\prime) \,] \,\rangle\, 
G(x-x^\prime) \, \Gamma_j^\nu(x^\prime\,y^\prime\,;z^\prime) \, G(y^\prime-y) \,] \nonumber\\\nonumber\\
&& -2 m^{-2} M \, \partial_z^0 \, [\,\delta(z-x) \,\gamma_0\,\gamma_5 \,\tau_i \,G(x-y)+\delta(z-y) \,G(x-y) \,\gamma_0 \gamma_5 \,\tau_i \,]
\end{eqnarray}
\\\hspace*{4.mm}
By the Fourier transform of Eq. (18) the relation of the current of the vertex function is obtained
\begin{eqnarray}
&&(p-q) \cdot \Gamma_i(p,q) \nonumber\\\nonumber\\
&&= -A^{-1} [\,(1+\gamma_5)\tau_i G(q)^{-1}-G(p)^{-1}(1-\gamma_5)\tau_i\,] -2 M \rho_5 \tau_i \qquad\qquad\nonumber\\\nonumber\\
&&-2 m^{-2} M A^{-1} g_{\mu 0} (p-q)^\mu \,[\,G(p)^{-1} \,\gamma_0 \gamma_5 \tau_i + \gamma_0 \gamma_5 \tau_i \,G(q)^{-1} \,]
\end{eqnarray}
\begin{eqnarray}
A \equiv 1 - (p-q)^2 / m^2
\end{eqnarray}
in the momentum space.
Under the on-shell condition of the incoming ($q$) and the outgoing ($p$) momenta of nucleon such as $\gamma\cdot p \rightarrow M$ and $\gamma\cdot q \rightarrow M$
Eq. (19) becomes $(p-q) \cdot \Gamma_i(p,q) \rightarrow -2 M \rho_5 \tau_i$ and the breaking term of the conservation of the current remains.
The term acts as the pole term of the vertex $\Gamma_i(p,q)$ and contributes to the interaction.
\\
\section*{\normalsize{3 \quad The effect of the pole term on the $\beta$-decay of neutron}}
\hspace*{4.mm}
The higher-order correction of the pole term in Eq. (19) by the pion propagator is essential because the term of the lowest-order cancels with the other term
as seen later.
In the present study the vertex correction is calculated by the perturbative expansion with the pseudovector coupling interaction lagrangian density
\begin{eqnarray}
L_{pv} = -\frac{f}{m_\pi} \bar{\psi}(x) \,\gamma_5 \gamma_\mu \vec{\tau} \,\psi(x) \,\partial^\mu \vec{\varphi}(x) 
\end{eqnarray}
where $f$ and $m_\pi$ are the coupling constant and the mass of pion respectively.
\\\hspace*{4.mm}
Recently we have suggested that the pion-nucleon-nucleon three point vertex has the non-perturbative term 
\begin{eqnarray}
\Gamma(p,q) = \gamma_5 \gamma \cdot (p-q) +G(p)^{-1} \,\gamma_5 +\gamma_5 \, G(q)^{-1}
\end{eqnarray}
in which the perturbative term is represented by the lowest-order approximation herein.
Leaving the self-energy out the vertex is in agreement with the pseudoscalar coupling irrespective of the on-shell condition.
\\\hspace*{4.mm}
The exact nucleon propagator $G(p)$ is expressed as 
\begin{eqnarray}
G(p) = (\gamma \cdot p-M-\Sigma(p))^{-1} 
\end{eqnarray}
along with the self-energy $\Sigma(p)$.
Because of the non-perturbative term in Eq. (22) the quadratic divergent term in the numerator of the fraction cancels out the denominator.
The convergent result of $\Sigma(p)$ is obtained
\begin{eqnarray}
\Sigma(p) = M c_1(p^2) - \gamma \cdot p \, c_2(p^2)
\end{eqnarray}
in terms of the coefficients $c_i(p^2)$ ($i$=1,2) as a function of $p^2$.
These are expanded around the on-shell point $p^2 = M^2$.
Particularly the value of the zeroth order $c \equiv c_1^{(0)}(M^2) = c_2^{(0)}(M^2)$ is necessary to examine the process of the $\beta$ decay.
The higher-order coefficients are dropped by the on-shell condition for the momenta of the initial neutron and the final proton.  
\\\hspace*{4.mm}
Using two quantities to characterize the correction that is the self-energy $c$ and the coefficient of the pole term $a_{n}$ 
the three-point vertex part $\Gamma_i(p,q)$ is expressed as follows
\begin{eqnarray}
\Gamma_i(p,q) \rightarrow A^{-1} (1+c) \,[\, \gamma (1-\gamma_5) + 2 M (1+c)^{-1} \hat{c}\, (p-q)^{-2} (p-q) \gamma_5 \,]\,\tau_i \nonumber\\\nonumber\\
+ \,\Gamma^{\prime}_i(p,q) \qquad\qquad\qquad
\end{eqnarray}
\begin{eqnarray}
\hat{c} \equiv c+(A^{-1}- \sum_{n=0}^{\infty} a_{n})A
\end{eqnarray}
under the assumption that $\Gamma_i(p,q)$ is sandwitched by the Dirac spinors such as $\bar{u}^{(s^\prime)}(p) \Gamma_i(p,q) u^{(s)}(q)$.
The $\Gamma_i^\prime(p,q)$ represents the term for which the relation $(p-q) \Gamma_i^\prime(p,q) =0$ exists irrespective of the on-shell condition.
At the moment it is not included since the term is related to the magnetism of the vector current and the part of the axial-vector current is independent of 
the strong interaction by the invariance of the charge conjugation \cite{Lurie}.
When the mass of the vector boson is much larger than the momentum transfer ($m^2 \gg (p-q)^2$), then $A \approx 1$ 
and the relation in Eq. (26) is substituted to $\hat{c} = c -\sum_{n=1} a_{n}$ in good approximation.
Here the subscript of the coefficient $a_n$ in the pole term specifies the number of the propagator of pion.
The lowest-order $a_0 \,$(=1) does not contribute to the pole term.
\\\hspace*{4.mm}
The process mediated by the pion propagator is a part of $a_1$ and it plays a prominent role.
Since the diagram is related to the decay of the $\pi^{-}$ it is free from the result of the perturbative calculation
and the strength of the process depends on the decay constant in the model hamiltonian.
To calculate the correction on the $\beta$-decay we treat the main part of $a_1$ in a particular way different from the other part of the pole term. 
The momentum transfer is $(p-q)^2 \approx 0$ 
and the divergence of the current is undetermined as $\sim (p-q)^2/((p-q)^2 - m_\pi^2) \rightarrow 0$ unless the mass of the pion is neglected
because of the factor arising from the loop integral.
\\\hspace*{4.mm}
The effect of the pion propagator is important 
for the zeroth order of $\rho_5$ given as $\rho_5^{(0)} = a_0 \gamma_5$ is eliminated from the pole term.
Then our interest is to calculate the coefficient of the second-order $a_1 \propto (\frac{f}{m_\pi})^2$
at the limit $(p-q)^2 \rightarrow 0$ of the on-shell nucleons.
The coefficient $a_1$ of the vertex correction $\rho_5^{(2)} = a_1 \gamma_5$ is given in the form of the integral as 
\begin{eqnarray}
\rho_5^{(2)} = -(\frac{f}{m_\pi})^2 \int \frac{d^4 k}{i (4 \pi)^4} \Delta(k) \Gamma(p,p-k) G(p-k) \gamma_5 G(q-k) \Gamma(q-k,q)
\end{eqnarray}
with respect to the momentum $k$ of the virtual pion.
To make the evaluation of the integral in Eq. (27) tractable 
the pion propagator is approximated by the free one such as $\Delta(k) \approx \Delta_0(k) = (k^2 - m_\pi^2)^{-1}$
and the self-energy of the nucleon propagator is set equal to zero ($G \approx G_0$). 
The vertex part is simplified by neglecting the non-perturbative term ($\Gamma(p,q) \approx \gamma_5 \gamma \cdot (p-q)$). 
The integral is performed by using the dimensional reguralization method and it yields as
\begin{eqnarray}
a_1 = \frac{G_\pi^2}{(4 \pi)^2} \, ( 2 D -\frac{7}{6}-\,{\rm log} \frac{M}{m_\pi}) + O(m_\pi^2/M^2)
\end{eqnarray}
\begin{eqnarray}
D \equiv \frac{2}{\epsilon}+1-\gamma-{\rm log} \frac{m_\pi^2}{4 \pi \mu^2}
\end{eqnarray}
where $G_\pi \equiv 2 M f / m_\pi$ and $\gamma = 0.577\cdots$ is the Euler's constant.
The parameters $\epsilon$ and $\mu$ are ascribed to the shift of the dimension as ${\rm 4} \rightarrow {\rm 4}-\epsilon$.
The calculation of the form factor is useful to remove the divergence in $a_1$ without preparing the renormalized functions \cite{Kinpara}.
The relation $D=0$ has been used to obtain the finite value of the $\gamma$-N-N vertex as a function of the momentum transfer $Q^2$. 
Subtracting the divergent part by the above relation it yields $a_1 = -3.56085$ not including the term of the $O(m_\pi^2/M^2)$ order.
\\\hspace*{4.mm}
Our interest is the pole term in Eq. (25) which possibly has an effect on the ratio 
between the vector and the axial-vector coupling constants $\lambda \equiv g_A/g_V$.
Since there is not a mass term in the denominator the matrix element gives a non-zero value at the limit $(p-q)^2 \rightarrow 0$.
In the case of the $\beta$-decay the initial state of neutron stands with the spin directed to the $z$-axis and the final state of baryon is specified
by the proton momentum $\vec{p}$ and the spin.
The direction of momentum enters in the calculation through the parameter $\theta$ which represents the average value of the polar angle as 
${\rm cos} \,\theta =C/2$ using the asymmetry parameter of proton $C=-0.2377$ \cite{Schumann}.
\\\hspace*{4.mm}
At the limit $(p-q)^2 \rightarrow 0$ there exists the simple relation for the elements 
between the axial-vector part $\Gamma_A \equiv \bar{u}^{(s^\prime)}(p) \gamma \gamma_5 u^{(s)}(q)$ 
and the pseudoscalar part $\Gamma_P \equiv (p-q)^{-2}(p-q)\bar{u}^{(s^\prime)}(p) \gamma_5 u^{(s)}(q)$ as
\begin{eqnarray}
\Gamma_P \approx (2 M)^{-1} {\rm cos}^2 \theta \, \Gamma_A
\end{eqnarray}
making assumption of the nucleon spin unchanged by the decay ($s^\prime = s$).
Then $\Gamma(p,q)$ is modified as
\begin{eqnarray}
\Gamma(p,q) \rightarrow (1+c) (\gamma -\gamma \gamma_5 + \alpha \,\gamma \gamma_5)
\end{eqnarray}
\begin{eqnarray}
\alpha \equiv (1+c)^{-1} \, \hat{c} \; {\rm cos}^2 \theta
\end{eqnarray}
The size of $\alpha$ is connected with the assymmetry of the momentum of proton.
\\\hspace*{4.mm}
The pion effect of the pole term is interesting to study the $\beta$ decay in detail and search for the value of $c$ to construct the self-energy.
It has been known
that the ratio of the currents $\lambda \equiv g_A/g_V$ is related to the decay constant of the charged pion as 
\begin{eqnarray}
-\lambda = \frac{G_\pi F_\pi}{\sqrt{2} M}
\end{eqnarray}
where $G_\pi$ is same as the $\pi$-N coupling constant of the pseudoscalar pion \cite{Lurie}.
Making use of the value $f =1$ it yields $G_\pi = 13.4544$ appropriate to the calculations of the $\pi$-N elastic scattering.
\\\hspace*{4.mm}
The decay constant $F_\pi \,(> 0)$ is determined from the relation of the decay rate $\Gamma_\pi$ for the process $\pi^- \rightarrow \mu^- + \bar{\nu}_\mu$
in which the value of the lifetime of $\pi^{-}$ is given experimentally as $\Gamma_\pi^{-1} = 2.6033 \times 10^{-8} \,{\rm s}$.
Using the Fermi coupling constant $G_F = 1.16637 \times 10^{-5}\,{\rm GeV}^{-2}$ it results in $F_\pi = 0.12825 \,{\rm GeV}$.
These numerical values of the constants are applied to the relation in Eq. (33).
\\\hspace*{4.mm}
There is a difference of $- \lambda = 1.300$ from the current experimental value $-\lambda_{exp} = 1.2695 \pm 0.0029$ \cite{PDG} roughly a few percent.
We suggest modifying the $g_A /g_V$ ratio as $\lambda \rightarrow \lambda + \alpha$ to account for the $\beta$ decay correctly. 
The correction $\alpha$ depends on the self-energy and it is meaningful to examine the relation with the other phenomena.
Using the value $c = 2$ of the $n = 2$ model the corrected value results in $-(\lambda + \alpha) = 1.2734$.
Around the region $c$ moves $\alpha$ is positive 
and the improvement of the numerical value indicates the similarity to the electromagnetic property of the magnetic moment
in favor of the universality of the electromagnetic and the weak interactions rather than the $\pi$-N elastic scattering 
under the strong interaction in the low-energy region below the resonance.
\\
\section*{\normalsize{4 \quad Summary and remarks }}
\hspace*{4.mm}
Concerning the pole term of the vertex in the $\beta$-decay the lowest-order is suppressed by the cancellation with the other term 
and required to take into account the higher-order corrections by the pion propagator.
It is consistent with the hypothesis of the conserved vector current which is not renormalized by the pion propagators.
In spite that the second-order process mediated by pion is the part of the proper vertex 
the process is represented by the decay constant of the charged pion well 
and we have used the standard way of the partially conserved axial-vector current to give most amount of the ratio. 
The second-order result achieves to correct the excess of the main term by means of the asymmetry parameter of proton 
and the self-energy to determine the electromagnetic interaction.
The numerical result is changed by adding the inversion of the proton spin 
and then it is compensated by the shift of the parameter of the self-energy to the lower values.
\\
\hspace{4.mm}

\end{document}